\documentstyle{article}

\newcommand{\ba}{\begin{eqnarray}}
\newcommand{\ea}{\end{eqnarray}}

\begin{document}
\pagestyle{plain}

\title{A comparison between algebraic models of molecular spectroscopy}

\author{R. Bijker$^{1)}$, A. Frank$^{1,2)}$, R. Lemus$^{1)}$, 
J.M. Arias$^{3)}$ and F. P\'erez-Bernal$^{3)}$\\
\and
\begin{tabular}{rl}
$^{1)}$ & Instituto de Ciencias Nucleares, U.N.A.M.,\\
        & A.P. 70-543, 04510 M\'exico D.F., M\'exico\\
$^{2)}$ & Instituto de F\'{\i}sica, Laboratorio de Cuernavaca,\\
        & A.P. 139-B, Cuernavaca, Morelos, M\'exico\\
$^{3)}$ & Departamento de F\'{\i}sica At\'omica, Molecular y Nuclear,\\
        & Facultad de F\'{\i}sica, Universidad de Sevilla,\\
        & Apdo. 1065, 41080 Sevilla, Espa\~na
\end{tabular}}
\date{}
\maketitle

\begin{abstract}
We discuss a symmetry-adapted algebraic (or vibron) model for 
molecular spectroscopy. The model is formulated in terms 
of tensor operators under the molecular point group.  
In this way, we have identified interactions that are absent in 
previous versions of the vibron model, in which the Hamiltonian 
is expressed in terms of Casimir operators and their products.  
The inclusion of these new 
interactions leads to reliable spectroscopic predictions. 
As an example we study the vibrational excitations of the 
methane molecule, and compare our results with those obtained in 
other algebraic models.
\end{abstract}

\vspace{2cm}
\begin{center}
Invited talk at `Symmetries in Science X',\\
Bregenz, Austria, July 13-18, 1997\\
\end{center}

\clearpage
\section{Introduction}

The development and refinement of experimental techniques in 
high resolution spectroscopy has generated a wealth of new data 
on rovibrational spectra of polyatomic molecules. 
Highly symmetric molecules, such as tetrahedral XY$_4$ systems, 
form an ideal testing ground. On the one hand, the high degree of 
symmetry tends to reduce the complexity of the spectrum and on the 
other hand, the use of symmetry concepts and group theoretical 
techniques may help to interpret the data and eventually suggest 
new experiments \cite{WDC,Bobin}. A good example is provided by 
the methane molecule, for which there exists a large amount of 
information on vibrational energies.  

{\it Ab initio} calculations for rovibrational spectra of molecular 
systems attempt exact solutions of the Schr\"odinger equation. 
These calculations involve several configurations associated with the 
molecular electronic states and yield the force field constants  
\cite{Raynes,Lee} from which the spectrum can be generated \cite{Hecht}. 
For small molecules this procedure is still feasible, but this is in 
general not the case for polyatomic molecules, due to the large size 
of the configuration space. Despite the progress made in {\it ab initio} 
calculations, a direct comparison with experimental vibrational energies 
of methane still shows large deviations, 
especially for vibrational states with a higher number of quanta. 

An alternative method is provided by algebraic (or vibron) models 
(for a review see \cite{vibron,thebook}). The general method 
consists of two ingredients: (i) the introduction of $U(k+1)$ as 
the spectrum generating algebra for $k$ degrees of freedom, and 
(ii) for a system of bosonic degrees of freedom the states are assigned 
to the symmetric representation $[N]$ of $U(k+1)$. 

In its original formulation \cite{vibron1,vibron2}, rotations and 
vibrations were treated simultaneously in terms of coupled $U(4)$ 
algebras: ${\cal G}=U_1(4) \otimes U_2(4) \otimes \ldots$~, by 
introducing a $U(4)$ algebra for each bond ($k=3$). 
The electronic degrees of freedom can be included by introducing a 
unitary group for the electrons \cite{FIL}.
For polyatomic molecules it was found to be more convenient to first 
separate the rotations and vibrations and subsequently to treat the 
vibrations in terms of coupled $U(2)$ algebras \cite{Roos,IO}: 
${\cal G}=U_1(2) \otimes U_2(2) \otimes \ldots$~,
introducing a $U(2)$ algebra for each interatomic potential ($k=1$). 
In this version of the vibron model the calculation of matrix elements 
is greatly simplified. An additional advantage is that it is well-suited 
to incorporate the underlying discrete symmetries. 

In a different approach, it has been suggested to use a $U(k+1)$ model 
for the $k=3n-3$ rotational and vibrational degrees of freedom of a 
$n$-atomic molecule \cite{BDL}. This model has the advantage that it 
incorporates all rotations and vibrations and takes into account the 
relevant point group symmetry. However, for larger molecules the number  
of possible interactions and the size of the Hamiltonian matrices 
increase rapidly. A similar approach can be used for the vibrations 
only \cite{LM}.  

In this contribution, we discuss a symmetry-adapted version of the 
vibron model \cite{Be4,X3,ozono,metano} which is very well suited 
to describe the vibrations of polyatomic molecules, especially  
those with a high degree of symmetry. The method is based on 
a set of coupled $U(2)$ algebras, whose generators are projected 
on tensor operators under the molecular point group. 
In order to illustrate these ideas 
we first review the main ingredients of the $U(2)$ vibron model, 
its connection with the Morse oscillator and the harmonic limit.
Next we develop the formalism in more detail and take as an example 
the methane molecule which has tetrahedral symmetry. Wherever possible, 
we make a comparison between the present formulation and other 
algebraic models. 

\section{The U(2) vibron model}

The model is based on the isomorphism of the $U(2)$ Lie
algebra and the one-dimensional Morse oscillator, whose 
eigenstates can be associated with $U(2) \supset SO(2)$ states 
$|[N],m \rangle$ \cite{AGI}. 
The $U(2) \supset SO(2)$ algebra is generated by the set   
$\{ \hat G \} \equiv $  $ \{ \hat N, \, \hat J_{+}, 
\, \hat J_{-}, \, \hat J_{0} \}$, which satisfies the commutation 
relations 
\ba 
\left[ \hat J_0, \hat J_{\pm} \right] \;=\; \pm \hat J_{\pm} ~, 
\hspace{1cm} \left[ \hat J_+, \hat J_- \right] \;=\; 2\hat J_0 ~, 
\hspace{1cm} \left[ \hat N, \hat J_{\mu} \right] \;=\; 0 ~,
\ea 
with $\mu=\pm,0$. For the symmetric irreducible representation
$[N]$ of $U(2)$, the Casimir operator is given by 
$\vec{J}^{\, 2} = \hat N(\hat N+2)/4$, 
from which follows the identification $j=N/2$. The $SO(2)$
label is denoted by $m$. 
The Hamiltonian 
\ba 
\hat H_M \;=\;  \frac{A}{2} 
\left( \hat J_- \hat J_+ + \hat J_+ \hat J_- \right) 
\;=\;  A \left( \vec{J}^2 - \hat J_0^2 \right) ~. 
\label{hmorse}
\ea
corresponds to the Morse oscillator with energies 
\ba
E_M \;=\; A \left[ j(j+1)-m^2 \right] 
\;=\; AN \left[ (v+\frac{1}{2}) - \frac{v^2}{N} \right] ~, 
\label{emorse}
\ea
where the label $v=j-m$ denotes the number of quanta in the oscillator. 
The Morse eigenstates are denoted by $| [N],v \rangle$ with 
$v=0,1,\ldots,[N/2]$. The first term in $E_M$ is the harmonic 
contribution, whereas the second term represents the anharmonicity 
which vanishes in the large $N$ limit. 

The concept of the harmonic limit provides a link with a 
geometrical picture, and hence can be used to compare 
various models of molecular structure.
Here we apply this procedure for the $U(2)$ vibron model. 
The action of $\hat J_{\mu}$ on the Morse eigenstates is 
\ba 
\hat J_+ \, | [N],v \rangle &=& \sqrt{ (N-v+1)v } \, | [N],v-1 \rangle ~,
\nonumber\\
\hat J_- \, | [N],v \rangle &=& \sqrt{ (N-v)(v+1) } \, | [N],v+1 \rangle ~.
\nonumber\\
2\hat J_0 \, | [N],v \rangle &=& (N-2v) \, | [N],v \rangle ~.
\ea 
Next we scale the generators $\hat J_{\mu}$ appropriately 
and take the limit $v/N \rightarrow 0$
\ba
\lim_{v/N \rightarrow 0} \frac{\hat J_+}{\sqrt{N}} \, | [N],v \rangle 
&=& \sqrt{v} \, | [N],v-1 \rangle \;\equiv\; a \, | [N],v \rangle ~,
\nonumber\\
\lim_{v/N \rightarrow 0} \frac{\hat J_-}{\sqrt{N}} \, | [N],v \rangle 
&=& \sqrt{v+1} \, | [N],v+1 \rangle 
\;\equiv\; a^{\dagger} \, | [N],v \rangle ~.
\nonumber\\
\lim_{v/N \rightarrow 0} \frac{2\hat J_0}{N} \, | [N],v \rangle
&=& (1-\frac{2v}{N}) \, | [N],v \rangle 
\;\equiv\; (1-\frac{2}{N} a^{\dagger}a) \, | [N],v \rangle ~,
\nonumber\\
\frac{\hat N}{N} \, | [N],v \rangle
&=& | [N],v \rangle ~.
\ea
In the harmonic limit ($v/N \rightarrow 0$), the $U(2)$ algebra 
contracts to the Heisenberg-Weyl algebra 
which is generated by $a^{\dagger} a$, $a^{\dagger}$, $a$ and 1.
For example, we have   
\ba
\frac{1}{N} \left[ \hat J_+,\hat J_- \right] \;=\; 
\frac{1}{N} 2 \hat J_0 \hspace{1cm} &\rightarrow& \hspace{1cm} 
\left[ a,a^{\dagger} \right] \;=\; 1 ~.
\ea
The other commutation relations can be treated similarly.
The application of the 
harmonic limit of the Morse Hamiltonian of Eq.~(\ref{hmorse}) gives 
\ba
\lim_{N \rightarrow \infty} \frac{1}{N} \hat H_M &=&  
\lim_{N \rightarrow \infty} \frac{A}{2N} 
\left( \hat J_- \hat J_+ + \hat J_+ \hat J_- \right) 
\nonumber\\
&=& \frac{A}{2} ( a^{\dagger} a + a a^{\dagger}) 
\;=\; A ( a^{\dagger} a + \frac{1}{2} ) ~,
\ea
with energies
\ba 
\lim_{N \rightarrow \infty} \frac{1}{N} \hat E_M &=&  A(v+\frac{1}{2}) ~,
\ea
in agreement with the large $N$ limit of Eq.~(\ref{emorse}).

\section{Symmetry-adapted algebraic model}

The $U(2)$ model described above was introduced to treat the 
stretching vibrations of diatomic molecules \cite{Roos}. 
For polyatomic molecules
it was suggested to treat the vibrational excitations in terms of 
coupled $U(2)$ algebras. This formulation was found to be very well 
suited to incorporate the underlying discrete symmetries \cite{IO}. 
In particular, invariant interactions under the point group were
constructed by applying proyection techniques on an expansion of 
the Hamiltonian in terms of Casimir invariants. 
In this section, we apply this process of symmetry adaptation to 
the generators of the $U(2)$ algebras themselves, rather than to 
the Casimir operators. This procedure leads to new interaction 
terms. 

We illustrate the method by an application to the stretching 
and bending vibrations of methane.
In the present approach, we associate a $U(2)$ algebra with 
each relevant interatomic interaction. For the CH$_4$ molecule we have 
four $U(2)$ algebras corresponding to the C-H interactions and six more 
representing the H-H couplings. 
The molecular dynamical group is then given by the product 
${\cal G}=U_1 (2) \otimes \ldots \otimes U_{10}(2)$,  where 
each $U_i(2)$ algebra is generated by the set 
$\{ \hat G_i \} \equiv \{ \hat N_i, \, \hat J_{+,i}, 
\, \hat J_{-,i}, \, \hat J_{0,i} \}$, which satisfies the commutation 
relations 
\ba
\, [ \hat J_{0,i}, \hat J_{\pm,i}] \;=\; \pm \hat J_{\pm,i} ~,
\hspace{1cm} 
\, [ \hat J_{+,i}, \hat J_{-,i}] \;=\; 2 \hat J_{0,i} ~,
\hspace{1cm} 
\, [ \hat N_i, \hat J_{\mu,i}] \;=\; 0 ~, \label{jmui}
\ea
with $\mu=\pm,0$. The labeling is such that $i=1,\ldots,4$ correspond to 
the C-H couplings while the other values of $i$ are associated with 
H-H interactions \cite{LF}. Here $\hat N_i$ is the $i$-th number 
operator. All physical operators are 
expressed in terms of the generators $\{ \hat G_i \}$, and hence 
commute with the number operators $\hat N_i$. 
For the CH$_4$ molecule there are two different boson 
numbers, $N_s$ for the C-H couplings and $N_b$ for the H-H couplings,
which correspond to the stretching and bending modes, respectively.

The tetrahedral symmetry of methane is taken into account by projecting 
the local operators $\{ \hat G_i \}$, which act on bond $i$, on the 
irreducible representations $\Gamma$ of the tetrahedral group 
${\cal T}_d$. For the $\hat J_{\mu,i}$ generators of Eq.~(\ref{jmui})
we obtain the ${\cal T}_d$ tensors 
\ba
\hat T^{\Gamma_x}_{\mu,\gamma} &=& 
\sum_{i=1}^{10} \, \alpha^{\Gamma_x}_{\gamma,i} \, \hat J_{\mu,i} ~,
\label{alpha}
\ea
where  $\gamma$ denotes the component of $\Gamma$, 
and the label $x$ refers to stretching ($s$) or bending ($b$). 
The expansion coefficients are the same as those given in \cite{LF} for 
the one-phonon wave functions. 
The algebraic Hamiltonian is now constructed by repeated couplings 
of these tensors to a total symmetry $A_1$. 

The methane molecule has nine vibrational degrees of freedom. Four of 
them correspond to the fundamental stretching modes ($A_1 \oplus F_2$) 
and the other five to the fundamental bending modes ($E \oplus F_2$) 
\cite{Herzberg}. A convenient labeling for the vibrational levels of 
CH$_4$ is provided by $(\nu_1,\nu_2^{l_2},\nu_3^{l_3},\nu_4^{l_4})$, 
where $\nu_1$, $\nu_2$, $\nu_3$ and 
$\nu_4$ denote the number of quanta in the $A_{1,s}$, $E_b$, 
$F_{2,s}$ and $F_{2,b}$ modes, respectively. The labels $l_i$ are 
related to the vibrational angular momentum associated with degenerate 
vibrations. The allowed values are $l_i=\nu_i,\nu_i-2,\ldots,1$ or 0 
for $\nu_i$ odd or even \cite{Herzberg}.
The projected tensors of Eq.~(\ref{alpha}) 
correspond to ten degrees of freedom, four of which 
($A_1 \oplus F_2$) are related to stretching modes and six 
($A_1 \oplus E \oplus F_2$) to the bendings. Consequently we can 
identify the tensor $\hat{T}^{A_{1,b}}_{\mu,1}$ as the 
operator associated to a spurious mode. 
This identification makes it possible to eliminate the spurious states 
{\em exactly}. This is achieved by (i) ignoring the 
$\hat{T}^{A_{1,b}}_{\mu,1}$ tensor in the construction of the Hamiltonian, 
and (ii) diagonalizing this Hamiltonian in a symmetry-adapted basis from 
which the spurious mode has been removed following the procedure of 
\cite{LF,comment}. 

\subsection{Zeroth order Hamiltonian}

According to the above procedure, we now construct the ${\cal T}_d$ 
invariant interactions that are at most quadratic in the generators 
and conserve the total number of quanta 
\ba
\hat{\cal H}_{\Gamma_x} &=& \frac{1}{2N_{x}} \sum_{\gamma} \left( 
  \hat T^{\Gamma_x}_{-,\gamma} \, \hat T^{\Gamma_x}_{+,\gamma}
+ \hat T^{\Gamma_x}_{+,\gamma} \, \hat T^{\Gamma_x}_{-,\gamma} 
\right) ~,
\nonumber\\
\hat{\cal V}_{\Gamma_x} &=& \frac{1}{N_{x}} \sum_{\gamma} \,
\hat T^{\Gamma_x}_{0,\gamma} \, \hat T^{\Gamma_x}_{0,\gamma} ~.
\label{hv}
\ea
Here $\Gamma=A_1$, $F_2$ for the stretching vibrations ($x=s$) and 
$\Gamma=E$, $F_2$ for the bending vibrations ($x=b$). In addition 
to Eq.~(\ref{hv}), there are two stretching-bending interactions
\ba
\hat{\cal H}_{sb} &=& \frac{1}{2\sqrt{N_sN_b}} \sum_{\gamma} \left( 
  \hat T^{F_{2,s}}_{-,\gamma} \, \hat T^{F_{2,b}}_{+,\gamma}
+ \hat T^{F_{2,s}}_{+,\gamma} \, \hat T^{F_{2,b}}_{-,\gamma} \right) ~,
\nonumber\\
\hat{\cal V}_{sb} &=& \frac{1}{\sqrt{N_sN_b}} \sum_{\gamma} \, 
\hat T^{F_{2,s}}_{0,\gamma} \, \hat T^{F_{2,b}}_{0,\gamma} ~.
\label{hvsb}
\ea
The zeroth order vibrational Hamiltonian is now written as
\ba
\hat H_0 &=& \omega_1 \, \hat{\cal H}_{A_{1,s}} 
   + \omega_2 \, \hat{\cal H}_{E_b} 
   + \omega_3 \, \hat{\cal H}_{F_{2,s}} 
   + \omega_4 \, \hat{\cal H}_{F_{2,b}} 
   + \omega_{34} \, \hat{\cal H}_{sb} 
\nonumber\\
&& + \alpha_2 \, \hat{\cal V}_{E_b}  
   + \alpha_3 \, \hat{\cal V}_{F_{2,s}} 
   + \alpha_4 \, \hat{\cal V}_{F_{2,b}} 
   + \alpha_{34} \, \hat{\cal V}_{sb} ~. \label{h0}
\ea
The interaction $\hat{\cal V}_{A_{1,s}}$ has not been included, 
since the linear combination 
\ba
\sum_{\Gamma=A_1,F_2} (\hat{\cal H}_{\Gamma_s} + \hat{\cal V}_{\Gamma_s}) 
&=& \frac{1}{4N_s} \sum_{i=1}^4 \hat N_i(\hat N_i+2) ~,
\ea
corresponds to the constant contribution $N_s+2$ to the energies. 
Similarly, for the bending vibrations the sum of the terms  
\ba 
\sum_{\Gamma=A_1,E,F_2} (\hat{\cal H}_{\Gamma_b} + \hat{\cal V}_{\Gamma_b}) 
&=& \frac{1}{4N_b} \sum_{i=5}^{10} \hat N_i(\hat N_i+2) ~,
\ea
corresponds to a constant $3(N_b+2)/2$. However, in this case 
the interactions $\hat{\cal H}_{A_{1,b}}$ and $\hat{\cal V}_{A_{1,b}}$ 
have already been excluded in order to remove 
the spurious contributions from the Hamiltonian. 

The Hamiltion of Eq.~(\ref{h0}) is equivalent to an expansion 
in terms of Casimir operators. It has the advantage, though, that 
the spurious contributions have been eliminated from the outset. 
A comparison with the Hamiltonian of \cite{LF} yields three 
conditions on their parameters 
\ba 
A_5 + 2 B_{5,10} + 8 B_{5,6} &=& 0 ~, 
\nonumber\\ 
B_{1,5} + B_{1,8} &=& 0 ~, 
\nonumber\\ 
\lambda_{1,5} + \lambda_{1,8} &=& 0 ~. 
\ea
The first condition eliminates the spurious interaction from 
the bending Hamiltonian of \cite{LF}, whereas the latter two eliminate the 
spurious contributions from the stretching-bending interactions. 
We note that the condition on the Hamiltonian that was used in 
\cite{LF} to exclude the spurious terms, does not automatically 
hold for states with higher number of quanta, nor does it remove 
all spurious contributions.  

\subsection{Harmonic limit}

In the harmonic limit the interaction terms of Eq.~(\ref{h0}) 
have a particularly simple form, which can be 
directly related to configuration space interactions 
\ba
\lim_{N_{x} \rightarrow \infty} \, \hat{\cal H}_{\Gamma_x} 
&=& \frac{1}{2} \sum_{\gamma} \left( 
  a^{\Gamma_x \, \dagger}_{\gamma} \, a^{\Gamma_x}_{\gamma}
+ a^{\Gamma_x}_{\gamma} \, a^{\Gamma_x \, \dagger}_{\gamma} \right) ~,
\nonumber\\
\lim_{N_{x} \rightarrow \infty} \, \hat{\cal V}_{\Gamma_x} &=& 0 ~,
\nonumber\\
\lim_{N_s, N_b \rightarrow \infty} \, \hat{\cal H}_{sb} 
&=& \frac{1}{2} \sum_{\gamma} \left( 
  a^{F_{2,s} \, \dagger}_{\gamma} \, a^{F_{2,b}}_{\gamma}
+ a^{F_{2,s}}_{\gamma} \, a^{F_{2,b} \, \dagger}_{\gamma} \right) ~,
\nonumber\\
\lim_{N_s, N_b \rightarrow \infty} \, \hat{\cal V}_{sb} &=& 0 ~.
\label{harlim}
\ea
Here the operators $a^{\Gamma_x}_{\gamma}$ are given in 
terms of the local boson operators $a_i$ through 
the coefficients $\alpha^{\Gamma_x}_{\gamma,i}$ given in 
Eq.~(\ref{alpha}) 
\ba
a^{\Gamma_x}_{\gamma} &=& 
\sum_{i=1}^{10} \, \alpha^{\Gamma_x}_{\gamma,i} \, a_i ~,
\ea
with a similar relation for the creation operators. From 
Eq.~(\ref{harlim}) the physical interpretation of the interactions 
is immediate. The $\hat{\cal H}_{\Gamma_x}$ terms represent the 
anharmonic counterpart of the harmonic interactions, while 
the $\hat{\cal V}_{\Gamma_x}$ terms are purely anharmonic 
whose contribution to the excitation energies vanishes 
in the harmonic limit. 

We note, that the recently introduced boson-realization model \cite{chinese} 
corresponds to the harmonic limit of the present approach, since 
it is formulated directly in terms of the boson creation 
and anniliation operators, $a^{\Gamma_x,\, \dagger}_{\gamma}$ and 
$a^{\Gamma_x}_{\gamma}$. The difference between the two 
lies in the anharmonic contributions which are implicit in the 
$U(2)$ approach, but which vanish in the harmonic limit \cite{comment}. 

\subsection{Higher order interactions}

The zeroth order Hamiltonian of Eq.~(\ref{h0}) is not sufficient to  
obtain a high-quality fit of the vibrations of methane. For example,  
the results presented in \cite{LF} were obtained by fitting 19 
vibrational energies with a r.m.s. deviation of 12.16 cm$^{-1}$. 
The boson-realization model of \cite{chinese} which, as was shown 
above, corresponds to the harmonic limit of the present approach 
was applied to  
the same 19 vibrations with a r.m.s. deviation of 11.61 cm$^{-1}$. 
We note, however, that the latter calculation includes some higher 
order interactions, without significantly improving the results. 

Several physically meaningful interaction terms that are essential 
for an improved fit are not present in Eq.~(\ref{h0}). 
They arise in the present model as higher order interactions. 
Products of $\hat{\cal H}_i$ and $\hat{\cal V}_j$ 
\ba
\hat{\cal H}_i \hat{\cal H}_j ~, \hspace{1cm} 
\hat{\cal V}_i \hat{\cal V}_j ~, \hspace{1cm} 
\hat{\cal H}_i \hat{\cal V}_j ~, 
\ea
are equivalent to an expansion in powers of Casimir operators. 
These terms only involve intermediate couplings with 
$\Gamma=A_1$ symmetry, since $\hat{\cal H}_i$ and $\hat{\cal V}_j$ 
themselves are scalars under the tetrahedral group.
However, there exist other interaction terms that involve intermediate 
couplings with $\Gamma=A_2$, $F_1$, $E$, $F_2$ symmetry. 
For example, the interactions   
\ba
g_{22} \, \hat l^{A_2} \, \hat l^{A_2} 
   + g_{33} \, \sum_{\gamma} \hat l^{F_1}_{s,\gamma} \, 
                             \hat l^{F_1}_{s,\gamma} 
   + g_{44} \, \sum_{\gamma} \hat l^{F_1}_{b,\gamma} \, 
                             \hat l^{F_1}_{b,\gamma} 
   + g_{34} \, \sum_{\gamma} \hat l^{F_1}_{s,\gamma} \, 
                             \hat l^{F_1}_{b,\gamma} ~,
\ea
with
\ba 
\hat l^{A_2} &=& -i \, \sqrt{2} \frac{1}{N_b} 
[ \hat T^{E_b}_{-} \times \hat T^{E_b}_{+} ]^{A_2} ~,
\nonumber\\
\hat l^{F_1}_{x,\gamma} &=& +i \, \sqrt{2} \frac{1}{N_x} 
[ \hat T^{F_{2,x}}_{-} \times \hat T^{F_{2,x}}_{+} ]^{F_1}_{\gamma} ~.
\label{vibang}
\ea
split levels with the same $(\nu_1,\nu_2,\nu_3,\nu_4)$, 
but with different $l_2$, $l_3$ and/or $l_4$. 
The square brackets in Eq.~(\ref{vibang}) denote the tensor couplings  
under the point group ${\cal T}_d$. 
Similarly, all higher order terms and anharmonicities can be 
constructed in a systematic way. Each one of the interaction terms 
has a direct physical interpretation and a specific action on the 
various modes. 

For the study of the vibrational excitations of methane we propose 
the following ${\cal T}_d$ invariant Hamiltonian \cite{metano,SS9} 
\ba
\hat H &=& \omega_1 \, \hat{\cal H}_{A_{1,s}} 
         + \omega_2 \, \hat{\cal H}_{E_b} 
         + \omega_3 \, \hat{\cal H}_{F_{2,s}} 
         + \omega_4 \, \hat{\cal H}_{F_{2,b}} 
         + \alpha_3 \, \hat{\cal V}_{F_{2,s}} 
\nonumber\\
&& + X_{11} \left( \hat{\cal H}_{A_{1,s}} \right)^2
   + X_{22} \left( \hat{\cal H}_{E_{b}  } \right)^2
   + X_{33} \left( \hat{\cal H}_{F_{2,s}} \right)^2
   + X_{44} \left( \hat{\cal H}_{F_{2,b}} \right)^2
\nonumber\\
&& + X_{12} \left( \hat{\cal H}_{A_{1,s}} \, \hat{\cal H}_{E_b    } \right)
   +  X_{14} \left( \hat{\cal H}_{A_{1,s}} \, \hat{\cal H}_{F_{2,b}} \right)
\nonumber\\
&& + X_{23} \left( \hat{\cal H}_{E_b    } \, \hat{\cal H}_{F_{2,s}} \right) 
   + X_{24} \left( \hat{\cal H}_{E_b    } \, \hat{\cal H}_{F_{2,b}} \right) 
   + X_{34} \left( \hat{\cal H}_{F_{2,s}} \, \hat{\cal H}_{F_{2,b}} \right) 
\nonumber\\
&& + g_{22} \, \hat l^{A_2} \, \hat l^{A_2} 
   + g_{33} \, \sum_{\gamma} \hat l^{F_1}_{s,\gamma} \, 
                             \hat l^{F_1}_{s,\gamma} 
   + g_{44} \, \sum_{\gamma} \hat l^{F_1}_{b,\gamma} \, 
                             \hat l^{F_1}_{b,\gamma} 
   + g_{34} \, \sum_{\gamma} \hat l^{F_1}_{s,\gamma} \, 
                             \hat l^{F_1}_{b,\gamma} 
\nonumber\\
&& + t_{33} \, \hat{\cal O}_{ss}
   + t_{44} \, \hat{\cal O}_{bb}
   + t_{34} \, \hat{\cal O}_{sb}
   + t_{23} \, \hat{\cal O}_{2s}
   + t_{24} \, \hat{\cal O}_{2b} ~. \label{hamilt}
\ea
The interpretation of the $\omega_i$ and $\alpha_3$ terms follows from 
Eq.~(\ref{harlim}). The $X_{ij}$ terms are quadratic in the operators 
$\hat{\cal H}_{\Gamma_x}$ and hence represent anharmonic vibrational 
interactions. The $g_{ij}$ terms are related to the vibrational angular 
momenta associated with the degenerate vibrations. As mentioned above,  
these interactions, which are fundamental to describe molecular systems with 
a high degree of symmetry, are absent in previous versions of the vibron   
model in which the interaction terms are expressed in terms of Casimir 
operators and products thereof. 
In the harmonic limit, the expectation 
value of the diagonal terms in Eq.~(\ref{hamilt}) leads to the familiar 
Dunham expansion \cite{Herzberg} 
\ba
\sum_i \omega_i \, (v_i + \frac{d_i}{2}) + \sum_{j \geq i} \sum_i 
X_{ij} \, (v_i + \frac{d_i}{2}) (v_j + \frac{d_j}{2})
+ \sum_{j \geq i} \sum_i g_{ij} \, l_i l_j ~. \label{Dunham}
\ea
Here $d_i$ is the degeneracy of the vibration. 
The $t_{ij}$ terms in Eq.~(\ref{hamilt}) 
give rise to further splittings of the vibrational levels 
$(\nu_1,\nu_2,\nu_3,\nu_4)$ into its possible sublevels. 
In the harmonic limit the $t_{ij}$ terms have the same interpretation 
as in \cite{Hecht}. The $\hat{\cal O}_{ss}$, $\hat{\cal O}_{bb}$ and 
$\hat{\cal O}_{sb}$ terms give rise to a splitting of the $E$ and $F_2$ 
vibrations belonging to the 
$(\nu_1,\nu_2^{l_2},\nu_3^{l_3},\nu_4^{l_4})=(0,0^0,2^2,0^0)$, 
$(0,0^0,0^0,2^2)$ and $(0,0^0,1^1,1^1)$ levels, respectively. 
Similarly, the $\hat{\cal O}_{2s}$ and $\hat{\cal O}_{2b}$ 
terms split the $F_1$ and $F_2$ vibrations belonging to the 
$(0,1^1,1^1,0^0)$ and $(0,1^1,0^0,1^1)$ overtones, respectively. 

\section{Results}  

The Hamiltonian of Eq.~(\ref{hamilt}) involves 23 interaction 
strengths and the two boson numbers, $N_s$ and $N_b$. The vibron 
number associated with the stretching vibrations is determined 
from the spectroscopic constants $\omega_e$ and $x_e \omega_e$ 
for the CH molecule to be $N_s=43$ \cite{LF}. The vibron number 
for the bending vibrations, which are far more harmonic than the 
stretching vibrations, is taken to be $N_b=150$. We have carried out 
a least-square fit to the vibrational spectrum of methane including 
44 energies. We find an overall fit to the observed levels 
with a r.m.s. deviation which is an order of magnitude better than 
in previous studies. While the r.m.s. deviations of \cite{LF} and 
\cite{chinese} are 12.16 and 11.61 cm$^{-1}$ for 19 energies, we find a 
r.m.s. of 1.16 cm$^{-1}$ for 44 energies. The values of the fitted
parameters as well as all predicted levels up to $V = 3$ can be found 
in \cite{metano,SS9}. 

The $\alpha_3$ term plays an important role in the calculation. 
It is completely anharmonic in origin and its contribution to the 
excitation energies vanishes in the harmonic limit. 
In order to address the importance of this term in  
Eq.~(\ref{hamilt}) we have carried out another calculation without 
this term. With one less interaction term the r.m.s. deviation 
increases from 1.16 to 4.48 cm$^{-1}$. This shows the importance of 
the term proportional to $\alpha_3$ to obtain an accurate description 
of the anharmonicities that are present in the data. 
The absence of the $\alpha_3$ term in the second calculation can 
only partially be compensated by  the anharmonicity 
constants $X_{ij}$. 

\section{Summary and conclusions}

In summary, we have discussed a symmetry-adapted algebraic model for 
molecular vibrations, in which the symmetry adaptation is applied at 
the level of the generators. This procedure has several interesting 
aspects: 
\begin{itemize} 

\item it provides a systematic procedure to construct all 
interaction terms up to a certain order, 

\item the harmonic limit gives a relation with configuration space 
interactions and Dunham expansions, 

\item the spurious states can be removed exactly. 

\end{itemize}
The application to the 44 observed vibrational excitations of methane 
gives a good overall fit with a r.m.s. deviation of 1.16 cm$^{-1}$ 
corresponding to an accuracy of $\sim 0.01 - 0.10 \, \%$,   
which can be considered of spectroscopic quality. 

It was pointed out that the ${\cal V}_{F_{2,s}}$ term in combination with 
the anharmonic effects in the other interaction terms plays a 
crucial role in obtaining a fit of this quality. Purely anharmonic terms 
of this sort arise naturally in the symmetry-adapted algebraic model, 
but vanish in the harmonic limit. Physically, these contributions arise
from the anharmonic character of the interatomic interactions, and seem 
to play an important role when dealing with molecular anharmonicities. 

We have established an explicit relation with the algebraic 
model of \cite{LF}, in which the Hamiltonian is expressed in terms of 
Casimir operators. A comparison between the two methods yields three 
constraints on the parameters, which remove the spurious components 
from the Hamiltonian of \cite{LF}. A comparison with the 
boson-realization model of \cite{chinese} shows that this model 
corresponds to the harmonic limit of the present approach. 

The predictability has been tested by systematically adding levels 
with higher number of quanta in the fitting procedure. The slow variation 
in the parameters shows that the model has a high degree of predictability. 
The application to methane \cite{metano} and to other molecules 
\cite{Be4,X3,ozono} 
suggest that the present model provides a numerically efficient tool 
to study molecular vibrations with high precision (r.m.s. deviations 
of $\sim$ 1 cm$^{-1}$). 

\section*{Acknowledgements}
This work was supported in part by the 
European Community under contract nr. CI1$^{\ast}$-CT94-0072, 
DGAPA-UNAM under project IN101997, 
and Spanish DGCYT under project PB95-0533.

\end{document}